# The Auger Effect and Ionization of Inner Atomic Shells

M.A. Kutlan

Institute for Particle & Nuclear Physics, Budapest, Hungary

kutlanma@gmail.com

The Auger effect from inner shells, which arises from resonance excitation of the valence electron of an atom by a weak electromagnetic wave, is discussed.

1. **INTRODUCTION**

The excitation of an electron shell arising in an atom can be removed in various ways, including through a nonradiative transition (Auger effect from inner shells). Various types of Auger processes arising from relaxation transitions in the electron shells of atoms have been closely studied both theoretically and experimentally (see, for example, the survey in [1] and [2]).

In the presence of an external electromagnetic wave, Auger transitions acquire several characteristic features. Experimental [3] and theoretical [4-6] studies have dealt with the effects of multiphoton excitation and ionization from the inner shells of many-electron atoms in the field of a strong electromagnetic wave. In this field, a large group of electrons of the outer shells participate in collective motion, and excite the core electrons through the Coulomb interaction.

Also of interest is a situation in which in the initial state of an atom located in the field of a weak electromagnetic wave, no vacancy is created in advance in its electron shell, and the quantum energy and wave intensity are insufficient for direct ionization from inner shells. In the case of resonance excitation of the valence electron by an external wave, relaxation to the ground state may be accompanied by spontaneous emission or by transfer of the wave energy to an electron of the inner shell. This transfer is due to the residual Coulomb interaction of the valence electron and an inner electron. Multiple repetition of this process may also lead to ionization of inner electrons.

In [7] expression was derived for the polarization potential, which acts on inner electrons in the course of resonance excitation of the valence electron. This potential was found by the psi-function formalism of a two-level system, and the probability of multiphoton ionization from the inner shells was estimated to be vanishingly small. For other aspects see [10-53].

## 2. THE PROBLEM

We consider an alkali metal atom in the field of a quasimonochromatic electromagnetic wave with a given law governing the switching-on of the electric field strength, E(t). The interaction of the valence electron with the wave will be described by the operator ($\hbar = c = 1$)

$$V(\mathbf{r},t) = -eE(t)(\mathbf{e}\mathbf{r})\cos\omega t \tag{1}$$

where $\mathbf{r}$ is the radius vector of the electron and $\mathbf{e}$ is the unit vector of polarization of the wave (below, a wave linearly polarized along the z axis will be considered: $\mathbf{e} = \mathbf{e}_z$ ).

We assume that when $t \to -\infty$, the field is switched off continuously, and when $t \to +\infty$, the field strength amplitude reaches a steady-state value $E_0$.

We consider the case of single-photon resonance, when for the valence electron the field frequency $\omega$ is close to the transition frequency: $\omega_{21} + \Delta$ and $|\Delta| \ll \omega_{21}, \omega$.

In addition, we assume that the frequency detuning $\Delta$ is smaller than the field width $\Gamma_f \sim d_{21}E_0$, where $d_{21}$ is the dipole matrix element of the transition in a two-level system.

The most appropriate method of describing the effect considered in this work is the approach using the density matrix formalism. In this approach, the Coulomb interaction of the valence electron with a core electron is described by the operator

$$V_e(\mathbf{r}_1,t) = e\int \rho(\mathbf{r}_2,t) \frac{e}{|\mathbf{r}_1-\mathbf{r}_2|}dV_2 = \int \rho_{21}(t)\psi_1^{(0)*}(\mathbf{r}_2) \frac{e^2}{|\mathbf{r}_1-\mathbf{r}_2|}\psi_2^{(0)}(\mathbf{r}_2)dV_2, \tag{2}$$

Wherep $\rho_{21}(t)$ is the nondiagonal element of the density matrix, calculated on the basis of the eigenfunctions $\psi_1^{(0)}$ and $\psi_2^{(0)}$ of the two-level system, and $\mathbf{r}_1$, and $\mathbf{r}_2$ are the radius vectors of the Auger electron and valence electron, respectively. We note that only the off-diagonal element of the density matrix, essential for the effect discussed, has been left standing in Eq. (2).

The equations describing the evolution of the density matrix contain the resonance interaction between the atom and the wave, together with the operator corresponding to dissipative processes [8].

The interaction (2) will be taken into account in accordance with perturbation theory. This approach implies the satisfaction of several criteria which must be met by the main parameters of the problem. Thus, the amplitude of the field strength $E_0$ of the wave must be large enough that the field width $\Gamma_f$ can be on the order of the spontaneous width $\Gamma_s$. As will be shown, the condition $\Gamma_f \sim \Gamma_s$ is optimal in terms of the probability of the Auger effect from the inner shell. On the other hand, the field $E_0$ must not be too high, so that the ionization probability of the valence is less than the probability of the Auger effect.

In addition, inclusion of the interaction (2) in accordance with perturbation theory assumes that the mixing amplitude of the resonance states in the external field is substantially greater than the mixing amplitude due to the residual Coulomb interaction with the Auger electron. Fulfillment of this condition imposes a certain upper bound on the allowed magnitude of the detuning $\Delta$. A criterion for the detuning $\Delta$ can be obtained, for example, from a comparative estimate of the diagonal corrections of the second approximation to the eigenfunctions of the system. The mixing amplitude of the resonance states by the external field is determined by the parameter $\alpha \approx (\Gamma_f / \Delta)^2$. A quantity having a similar meaning and arising from the residual Coulomb interaction of the valence electron and Auger electron is given by the parameter

$$\alpha_c = \int \frac{\left|\left\langle 1p \left| \frac{e^2}{|\mathbf{r}_1 - \mathbf{r}_2|} \right| 2i \right\rangle\right|^2}{\left[\varepsilon_p - \left(E_i^{(0)} + \omega\right)\right]^2} \frac{d\mathbf{p}}{(2\pi)^3}, \qquad (3)$$

where $\varepsilon_p = p^2 / 2m_e$ is the energy of the Auger electron in the continuous spectrum; $E_i^{(0)} = -I_0$ is the energy of this electron in the initial bound state ($I_0$ is the energy of ionization from an inner shell).

The matrix element in Eq. (3) is given by the expression

$$\left\langle 1p \left| \frac{e^2}{|\mathbf{r}_1 - \mathbf{r}_2|} \right| 2i \right\rangle = \int\int e^{-i\mathbf{p}\mathbf{r}_1} \psi_1^{(0)*}(\mathbf{r}_2) \frac{e^2}{|\mathbf{r}_1 - \mathbf{r}_2|} \psi_2^{(0)}(\mathbf{r}_2) \psi_i^{(0)}(\mathbf{r}_1) dV_1 dV_2, \qquad (4)$$

where the function $\psi_i^{(0)}(\mathbf{r}_1)$ describes the initial state of the Auger electron.

The calculation of the integral in Eq. (4) is appreciably simplified if one takes into consideration the comparative sizes of the outer shell and inner shells of the atom. The characteristic distances along $r_1$ and $r_2$ in Eq. (4) are determined by the radii of the corresponding shells: for r2, it is the radius of the valence shell of the atom, and for r,, the radius of the inner shell. As a rough estimate, one can assume that the radius of the ionizable shell is $r_s \sim a_0 Z_s Ry / I_0$, where $a_0 = \hbar^2 / m_e e^2$ is the first Bohr radius of the hydrogen atom; $Z_s$ is the effective charge of the core for the shell from which the Auger transition takes place; and $Re = m_e e^4 / 2\hbar^2 = 13.6 eV$.

It follows from the above estimate that for the electrons of inner shells $r_1 \ll r_2$, and in the dipole approximation the matrix element in Eq. (4) has the form

$$\left\langle 1p \left| \frac{e^2}{|\mathbf{r}_1 - \mathbf{r}_2|} \right| 2i \right\rangle = \int e^{-i\mathbf{p}\mathbf{r}_1} e\mathbf{r}_1 \psi_i^{(0)}(\mathbf{r}_1) dV_1 \int \psi_1^{(0)*}(\mathbf{r}_2) \nabla \left( \frac{e}{r_2} \right) \psi_2^{(0)}(\mathbf{r}_2) dV_2. \qquad (5)$$

Calculation using Ehrenfest's theorem and the conditions $I_0 \gg \omega$ of the integrals in Eq. (5) leads to the following estimate of the amplitude:

$$\alpha_c \approx \left( \frac{m_e \omega_{21}^2 a_0^2 \mathrm{Ry}}{I_0^2} \right)^2 \sim \left( \frac{I\,\mathrm{Ry}}{I_0^2} \right)^2,$$

where I is the binding energy of the valence electron.

From the estimates obtained, we have the following condition for the magnitude of the ratio of the field width to the frequency detuning

$$\Gamma_f / \Delta > I\,\mathrm{Ry} / I_0^2. \qquad (6)$$

In a real situation, it is not difficult to satisfy the inequality (6).

## 4. Discussion of Results

As an example, we consider the potassium atom ($5P_{1/2} \to 4S_{1/2}$ transition, $\hbar\omega = 3.07 eV$), for which the matrix elements of the corresponding dipole transition are known [9]. The intrinsic width of the $5P_{1/2}$ level is $\Gamma_s \approx 1.1 \times 10^{-7} eV$. The field width $\Gamma_f = (1/2)eE_0 z_{12}$ for the resonance transition turns out to be $\Gamma_f \approx 0.65 \times 10^{-8} E_0$ eV and becomes of order of the intrinsic width at low field strength, $E_0 \approx 17$ V/cm.

One more fact, which sets an additional lower bound on the allowable values of Eo, should be taken into consideration. The observation conditions should ensure that both the Doppler and collisional broadenings of the atomic levels be small compared to the field width: $\Gamma_D, \Gamma_C < \Gamma_f$. To decrease the effects associated with the Doppler width, the experiment should be performed with sufficiently well collimated atomic beams, the direction of the wave being transverse to the beam. At small beam divergence angles $\theta = d/2l$ (d being the channel diameter, and l the channel length), the transverse temperature of the atoms $T_\perp$ is related to the longitudinal temperature $T_\parallel = T$ as follows: $T_\perp \approx T\theta^2$. For $T = 10^{-3} K$, l = 1 m, and d = 1 cm, the Doppler width is $\Gamma_D \approx 5\times 10^{-8}$ eV, and thus, the field strength $E_0 \approx 17$ V/cm is sufficient to satisfy the inequality $\Gamma_f > \Gamma_D$.